\documentclass{rmf-d}
\usepackage{nopageno,rmfbib,multicol,times,epsf,amsmath,amssymb,cite}
\usepackage[T1]{fontenc} 
\usepackage[spanish, mexico]{babel}
\usepackage[]{caption2}
\usepackage{graphicx}
\usepackage{blindtext}
\usepackage[svgnames]{xcolor}
\usepackage[colorlinks,,citecolor=DarkGreen,linkcolor=DarkRed,urlcolor=DarkBlue]{hyperref}

\newcommand{\bs}{\boldsymbol}
\usepackage{slashed}

\def\rmfcornisa{ \hfill\ {\bf }
\hfill }

\begin{document}

%
%
\title{ Electric current from Schwinger's time-ordered propagator
}
\author{  Cristi\'an Villavicencio      }
\address{   Centro de Ciencias Exactas \& Departamento de Ciencias B\'asicas, Facultad de Ciencias, \\
Universidad del Bio-Bio,
Casilla 447, Chill\'an, Chile.    }
%
%
\maketitle
%
%
 \begin{abstract}
%
%
The quasistatic electric current density of fermions in the presence of an external electric field is determined through the utilization of a time-ordered Schwinger propagator.
The study encompasses the necessary conditions for establishing a well-defined time-ordered propagator within the Schwinger formalism, specifically concentrating on constant and uniform electromagnetic fields.
Within this theoretical framework, a chemical potential is introduced, resulting in non-zero values for the electric current and charge number density. Consequently, the dependence of electric conductivity on the strength of the electric field and the charge number density is investigated.

\vspace{1em}
\end{abstract}
\keys{ 
Schwinger's proper-time method; external electromagnetic field; electric conductivity.
}
\begin{multicols}{2}

Schwinger's proper time method has been a crucial tool for studying the effect of external electromagnetic fields in quantum field theory \cite{Schwinger:1951nm}.
In particular, the Green's function constructed by this method is often used in quantum field theory for the description of particles propagating under the influence of an external magnetic field, with the versatility of being expressed as an integral in proper time, in power series of Landau levels or expanded in magnetic field powers \cite{Chyi:1999fc,Kharzeev:2013jha}.

The effect of the electric field in the Schwinger formalism has been intensively studied due to the prediction of pair production in vacuum \cite{Sauter:1931zz,Schwinger:1951nm}.
But recently, there has been a growing interest in studying the effects of the electric field in the context of relativistic heavy-ion collisions, in particular for peripheral collisions or collisions with different ions such as Au-Cu \cite{Bzdak:2011yy,Deng:2012pc}.
This has motivated the study of quark and hadronic matter under external electric and magnetic fields by investigating the modifications of different parameters in systems in thermal equilibrium
\cite{Tavares:2018poq,Loewe:2022aaw,Endrodi:2022wym,Loewe:2022plp,Tavares:2023ybt} with particular interest in chiral restoration and deconfinement \cite{Cao:2015dya,Tavares:2019mvq,Ahmad:2020ifp, Loewe:2021mdo,Ahmad:2023mqg}, as well as anomalous transport phenomena \cite{PhysRevD.91.025011,PhysRevD.91.045001,Huang:2015oca,Copinger:2018ftr,Cao:2019hku} and also analogous systems in (2+1) dimensions \cite{Murguia:2009fs,Ayala:2010fm,Dudal:2021ret}.

The problem to be studied here concerns how to regularize the Schwinger's Green's function.
Integral expressions in the Schwinger propagator are commonly regulated by an infinitesimal constant that coincides with the time-ordered regulator in vacuum.
However, this is not the general case, and there are some details that are not usually considered in the treatment of this method, as was clearly detected in \cite{Tsamis:2000ah}, concluding that the Schwinger propagator is a Green's function, not a propagator itself, so some care must be taken when considering these Green's functions.

The correct integration of momentum in the complex plane is full of subtleties, particularly when considering medium effects such as temperature and chemical potential. This has been addressed, also including the effects of a constant and uniform external magnetic field
\cite{Shuryak:1980tp,PhysRevD.42.2881,Mizher:2013kza,Mizher:2018dtf}.
The presence of an external electric field introduces an additional problem, since it deals with time components, and therefore certain considerations must be taken into account when describing these systems in a real time formalism.
This was studied in \cite{Tsamis:2000ah} using light-cone coordinates, and in \cite{Copinger:2018ftr,Cao:2019hku} with in- and out-state expectation values.
The following discussion outlines the prerequisites for constructing Schwinger's time-ordered propagator in the presence of constant electric and magnetic background fields.
Next, the general regularization of the propagator in the Schwinger formalism is introduced.
The specific case of external uniform electric field is obtained, ending with the derivation and analysis of the electric charge density and electric current density for quasistatic motion.

\section{Time ordered propagator}

The usual prescription for regularizing a time-ordered propagator in momentum space is to shift the momentum $p^2\to p^2+i\epsilon$ in the Green's function.
This can be considered as a general statement if the Green's function is analytic near the poles of $p^2$, assuming poles on the real axis.
If the Lorentz symmetry is broken, the prescription must be generalized as $p_0\to p_0+i\epsilon p_0$, which must be taken into account, for example, if the chemical potential term is present \cite{Shuryak:1980tp}.
This is also general if the Green's function is analytic near the poles of $p_0$, assuming real poles.
In the case of the Schwinger propagator with a uniform external magnetic field and chemical potential, this regularization can be applied with some modifications in the integration in proper time \cite{PhysRevD.42.2881}.

The case of an external electric field is not as straightforward as that of a pure magnetic field, since its presence could introduce a non-local time component in Green's function through the Schwinger phase.
Therefore, we need to avoid this situation in order to have a well-defined time-ordered propagator.

Let us begin by considering the Green's function for fermions in configuration space
\begin{equation}
 G(x,x')=\langle x'|\frac{i}{\slashed{\Pi}-m}|x\rangle,
 \label{Eq.G}
\end{equation}
where the conjugate momenta operator is defined as $\Pi_\nu = \hat p_\nu+qA_\nu(\hat x)+\mu g_{\nu 0}$, where $q$ the charge ($q=-|e|$ in the case of the electron), $\hat x$ is the position operator, $\hat p$ is the momentum operator,  $A$ is the electromagnetic vector potential, $\mu$ is a chemical potential.

As it was remarked in \cite{Tsamis:2000ah}, the expression in Eq. (\ref{Eq.G}) is just a Green function.
However, we know that in the absence of external electromagnetic fields, or in the presence of an external uniform magnetic field, this Green's function becomes local in time coordinates, so that a well-defined Fourier transform can be performed, such that the prescription $p_0\to p_0+i\epsilon p_0$ makes sense.

If electric field is present, the locality argument is not in general valid, unless we consider the condition $\partial_0A(x)=0$.
In that case, $\hat p_0$ commutes with all other operators, and therefore Green's function becomes local in time coordinates.
\begin{equation}
G(x,x')\to G(x_0-x_0';\bs{x},\bs{x}').
\end{equation}
The time fourier transform can then be performed on the Green's function, or equivalently, a set of complete eigenstates of $\hat p_0$ can be introduced, leading to
\begin{equation}
\tilde G(p_0;\bs{x},\bs{x}')
 =\int d  t \,e^{ip_0 t}\,G(t;\bs{x},\bs{x}').
\end{equation}

We know that once Schwinger's procedure is applied, the Green's function can be written in terms of a local function multiplied by the Schwinger's phase, which is non-local:
\begin{equation}
 G(x,x')=e^{i\Phi(x,x')}S(x-x'),
 \label{eq.Shcwinger}
\end{equation}
where the Schwinger phase can be written as
\begin{equation}
 \Phi(x,x')=q\int_{x'}^{x}d\xi\cdot \tilde A(\xi),\qquad \xi=xt+x'(1-t),
\end{equation}
with the vector potential $\tilde A$ defined as $\tilde A_\nu = A_\nu + g_{\nu 0}\,\mu/q$.

Let's see in detail what happens with the Schwinger phase.
Since we are dealing with constant electromagnetic fields, considering the imposed condition that $\partial_0A(x)=0$, the components of the vector potential will be
\begin{equation}
 \tilde A_0(\bs{x}) = -E^ix_i +\mu/q, \qquad \tilde A_i(\bs{x})=\frac{1}{2}\epsilon_{ijk}B^jx^k,
\end{equation}
where the symmetric gauge for $A_i$ was considered.
The Schwinger phase is then
\begin{equation}
 \Phi(x,x')=(\mu -q\bs{E}\cdot\bs{\bar x})\,(x_0-x'_0)
 -[q\bs{B}\times\bs{\bar x}]\cdot(\bs{x}-\bs{x}'),
\label{eq.phase}
\end{equation}
where $\bs{\bar x}\equiv \frac{1}{2}(\bs{x}+\bs{x}')$ is the average position.
Now we can do the Fourier transform with respect to $x-x'$ obtaining as a result a Green's function in momentum space, but also as a function of the average position of the correlation.
The time-ordered, or Feynman propagator, is then
\begin{equation}
G_F(p;\bs{\bar x}) =\tilde S(P_0+i\epsilon p_0,\bs{P}),
\label{eq.Feynman}
\end{equation}
where $\tilde S$ is the Fourier transform of $S$ in Eq.\,(\ref{eq.Shcwinger}), with conjugate moments
\begin{equation}
 P_0=p_0+\mu-q\bs{E}\cdot\bs{\bar x},\qquad \bs{P} = \bs{p}+q\bs{B}\times\bs{\bar x}.
\end{equation}
We can see how the electric field immediately introduces a chemical potential term corresponding to the Coulomb potential for a constant electric field.
This is nothing new, since the chemical potential is commonly introduced as a zero component of the gauge electromagnetic field.
However, this chemical potential is position-dependent, so we have to see how to interpret it.

\section{Schwinger's Green function regularization}

The Schwinger Green function is commonly regularized by introducing an infinitesimal term to ensure that the integrals in proper time remain well-behaved and do not diverge.
The regulator will be incorporated by expressing the Green function in Eq.\,(\ref{Eq.G}) as
\begin{equation}
 G(x,x')=\langle x'|\frac{\slashed{\Pi}+m}{iH}|x\rangle,
 \label{eq.G-H}
\end{equation}
where the Hamiltonian is defined as
\begin{equation}
 H=-\slashed{\Pi}^2+m^2
 =-\Pi^2+\frac{q}{2}F_{\mu\nu}\sigma^{\mu\nu}+m^2.
\end{equation}
Subsequently, Eq\,(\ref{eq.G-H}) can be formulated in integral form
\begin{equation}
  G(x,x') =\int_0^\infty ds \langle x'|[\slashed{\Pi}+m]e^{-isH}|x\rangle.
  \label{eq.G-s1}
\end{equation}
To integrate, avoiding divergences, a regulator is introduced by shifting $H\to H-i\eta$.
If $\eta$ is positive, the convergence of the integration in $s$ is guaranteed.
Alternatively, if $\eta$ is negative, Eq.\,(\ref{eq.G-H}) can be formulated as
\begin{equation}
  G(x,x') =-\int_{-\infty}^0 ds \langle x'|[\slashed{\Pi}+m]e^{-isH}| x\rangle,
  \label{eq.G-s2}
\end{equation}
which is the same procedure as described in \cite{PhysRevD.42.2881}.
For the general case with an unknown regulator $\eta$, the Green's function can be written as
\begin{equation}
 G(x,x')= \int_{-\infty}^\infty  ds\,r_s(\eta)\,G_s(x,x')
  \label{eq.G-eta}
\end{equation}
with $r_s$ the general regulation function and $G_s$ the integrand of Eqs. (\ref{eq.G-s1}) and (\ref{eq.G-s2}) which can be written as
\begin{equation}
r_s(\eta) = e^{-s\eta} [ \theta(\eta)\theta(s)  -\theta(-\eta)\theta(-s)]
\end{equation}
\begin{align}
 G_s(x,x')=\langle x'(0)|[\slashed{\Pi}(0)-m]|x(s)\rangle.
 \label{eq.G(x,x';s)}
\end{align}
The Green's function can also be expressed in momentum space if we separate the correlation coordinates as $x=\bar x+\frac{1}{2}\Delta x$ and $x'=\bar x-\frac{1}{2}\Delta x$,  and then we apply the Fourier transform with respect to $\Delta x$:
\begin{equation}
 G(p;\bar x)= \int_{-\infty}^\infty  ds\,r_s(\eta)\,G_s(p;\bar x).
\end{equation}
The regulator $\eta$ is not restricted to being a constant, but can be a function of coordinates and/or momentum, so in principle it can be expressed differently in configuration space than in momentum space.
In fact, in the absence of external electromagnetic field, but with finite chemical potential, the regulator in momentum space is $\eta = \epsilon p_0(p_0+\mu)$.

\section{Electric field}

Let's now calculate the Schwinger's Green's function in the presence of an external uniform electric field along the third direction $\bs{E}=E\bs{\hat e}_3$.
In this case, calculating the Fourier transform of the integrand of the Green's function of Eq.\,(\ref{eq.G(x,x';s)}), the following expression is obtained:
\begin{equation}
\begin{split}
& G_s(p; z)
=\exp \left\{
is\left[\tau_s P_\parallel^2+p_\perp^2-m^2\right]\right\}
\\ & \quad\quad
\times\Big\{\slashed{P}_\parallel\left[
1+\tanh^2(qEs)-2\tanh(qEs)\gamma_0\gamma_3\right]
\\ & \qquad \qquad \qquad
+(\slashed{p}_\perp+m)\left[1-\tanh(qEs)\gamma_0\gamma_3\right]\Big\}
\end{split}
\label{eq.G_s}
\end{equation}
with
\begin{align}
 \tau_s &= \tanh(qEs)/qEs
 \label{eq.tau_s}\\
 P_\parallel &= (p^0+\mu-qEz,0,0,p^3)\\
 p_\perp &= (0,p^1,p^2,0)\\
 z &= \frac{1}{2}(x^3+x'^3),
\end{align}
and where the Minkowski metric convension used is $g_{\mu\nu}=g^\parallel_{\mu\nu}+g^\perp_{\mu\nu}$.
This result is directly derived from the general Schwinger Green's function \cite{Schwinger:1951nm,schwartz2014quantum}.

The Feynman propagator is then obtained by shifting $p_0\to p_0+i\epsilon p_0$.
The chemical potential plus the Coulomb potential now plays the role of an effective chemical potential
\begin{equation}
\tilde\mu=\mu-qEz,
\end{equation}
so, the following is obtained:
\begin{align}
 \eta & = \epsilon \tau_s\,2p_0(p_0+\tilde \mu).
\end{align}
Since $0\leq\tau_s\leq 1$, this function can be absorved in the infinitesimal parameter $\epsilon$.
This regulator is the same regulator with a chemical potential.
Finally, the Feynman propagator in momentum space is
\begin{equation}
 G_F(p;z) = \int_{-\infty}^\infty ds\,r_s\big(\epsilon p_0(p_0+\tilde \mu)\big)\,G_s(p; z),
 \label{eq.S_F}
\end{equation}

\section{Electric current and charge number density}

The expectation values of bilinear fermion operators are defined as
\begin{equation}
 \langle \bar\psi(x)\,\Gamma\,\psi(x) \rangle = -\textrm{tr}[\Gamma G(x,x)],
\end{equation}
where in this case $G$ refers to a time-ordered propagator that includes interactions.
The leading order expectation values for the electric current density operator and the charge number density operator are then
\begin{align}
 j&= q\,\langle \bar\psi\gamma^3\psi \rangle = -q\,\textrm{tr} \int\frac{d^4p}{(2\pi)^4} \, \gamma^3G_F(p;z),\\
 n &= \langle \psi^\dag\psi \rangle = -\textrm{tr} \int\frac{d^4p}{(2\pi)^4} \, \gamma^0 G_F(p;z),
\end{align}
respectively.
Both quantities depend now on the coordinate $z$.
which can be understood as the distance from the source of the constant magnetic field.
In the absence of a reference system, this interpretation is ambiguous, so the $z$ coordinate should be treated merely as a parameter that changes the value of the chemical potential.
However, as pointed out in \cite{Dittrich:2000zu}, under a gauge transformation, the chemical potential must be redefined to obtain the same physical system.
Then, unambiguously, we can calculate the electric current as a function of the number density and the electric field $j(n,E)$ parameterized by the effective chemical potential $\tilde \mu(n,E)$.

Using Eqs. (\ref{eq.G_s}) and (\ref{eq.S_F}), tracing over spin and integrating the spatial components of momentum,
the current density and charge number density are then
  \begin{align}
 j&= \frac{1}{2\pi^{5/2}}\int_{-\infty}^\infty ds\,f_s\,\frac{q^2E\sqrt{\tau_s}}{\sqrt{is}}
 \nonumber\\
 n &=\frac{1}{2\pi^{5/2}}\int_{-\infty}^\infty ds \,f_s\,\frac{\left[1+\tau_s^2(qEs)^2\right]}{2s\sqrt{\tau_s}\sqrt{is}}\nonumber
 \end{align}
 with
 \begin{equation}
  f_s \equiv \int_{-\infty}^\infty dp_0\, r_s\big(\epsilon p_0(p_0-\tilde\mu)\big)
ip_0 \, e^{ is\left[\tau_s\,p_0^2-m^2
 \right]},
 \label{eq:f_s}
 \end{equation}
and where $\tau_s$ was defined in Eq.\,(\ref{eq.tau_s}).
Note that the propagator in Eq.,\eqref{eq.G_s} involves only hyperbolic functions, in contrast to the case with an external magnetic field, where trigonometric functions appear in the denominator.
As a result, there are no poles here except at $s=0$, and therefore a Wick rotation is not required.
If a magnetic field were also considered, it might be more convenient to perform a Wick rotation. In that case, the electric field would be expressed in trigonometric functions, which would require a different treatment.
Integrating  Eq.\,\eqref{eq:f_s} in $p_0$ according to the different sign possibilities of the chemical potential and the proper time $s$, the electric current density and the charge number density becomes
\begin{align}
j = {}& \frac{\textrm{sign}(\tilde\mu)}{2\pi^{5/2}}\int_0^\infty  ds \,g_s\,
\frac{q^2E}{s^{3/2}\sqrt{\tau_s}},
\label{eq.j}
\\&\nonumber\\
n= {}& \frac{\textrm{sign}(\tilde\mu) }{2\pi^{5/2}}\int_0^\infty  ds\,
g_s\,\frac{1+\tau_s^2(qEs)^2}{2s^{5/2}\,\tau_s^{3/2}},
\label{eq.n}
\end{align}
with
\begin{equation}
 g_s\equiv
 \sin\left(sm^2-\frac{\pi}{4}\right)-\sin\left(s(m^2-\tau_s\tilde\mu^2)-\frac{\pi}{4}\right),
\end{equation}
where the phase factor $\pi/4$ comes from the $\sqrt{i}$ terms that appears in Ecs. (\ref{eq.j}) and (\ref{eq.n}).

\section{Results}

To obtain the electric current density as a function of the charge number density and the external electric field, we have to obtain the values of $n$ for different values of the effective chemical potential and the electric field.
The integrand of the number density $n$ in Eq.\,(\ref{eq.n}) is strongly divergent when $s \to 0$, making numerical integration difficult.
However, this can be handled by adding and subtracting its value in the limit $E \to 0$, but keeping the effective chemical potential finite
\begin{equation}
 n(\tilde \mu,E) = [n(\tilde \mu,E)-n(\tilde \mu,0)]+n_0(\tilde \mu),
\end{equation}
where $n_0$ is the well known value of the number density in vacuum at finite chemical potential
\begin{equation}
 n_0(\mu) =\frac{\epsilon(\mu)}{3\pi^2}\theta(|\mu|-m)(\mu^2-m^2)^{3/2}.
\end{equation}

It is more illustrative to show the electric conductivity $\sigma=j/E$.
The conductivity will be represented in one case as a function of the number density for a fixed value of the field $E$, and in the other case as a function of the external electric field for a fixed value of $n$.

\begin{figure}[H]
 \includegraphics[scale=0.65]{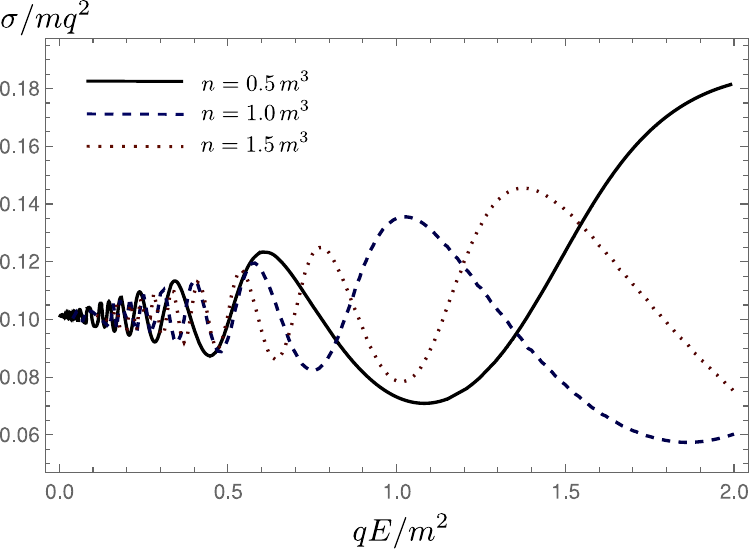}
 \caption{Electric conductivity normalized by $mq^2$ as a function of the external electric field number in units of the fermion mass for different values of the charge density.}
 \label{fig.nfixed}
\end{figure}

The case of electric current density as a function of external electric field for different values of charge number density is shown in Fig.\,\ref{fig.nfixed}.
For all number density values, the electrical conductivity starts from the same point at $E=0$, and then begins to oscillate as $E$ increases.
Let's call this initial value as $\sigma_0$ whose expression is
\begin{equation}
 \sigma_0 =\frac{q^2 m}{\pi^2}.
\end{equation}
Apparently, the only difference between the different values of the charge number density is a sort of phase in the oscillation around $\sigma_0$.
Although they look very similar, the oscillation amplitudes vary for the different cases, and their values are almost twice as large above $\sigma_0$ than below $\sigma_0$.

Caution is needed when interpreting this result for large electric fields, as the quasistatic motion assumption does not account for the induced magnetic field. In this context, the apparent oscillations may not necessarily have physical significance. These results are more reliable in the regime of weak electric fields.

\begin{figure}[H]
 \includegraphics[scale=0.65]{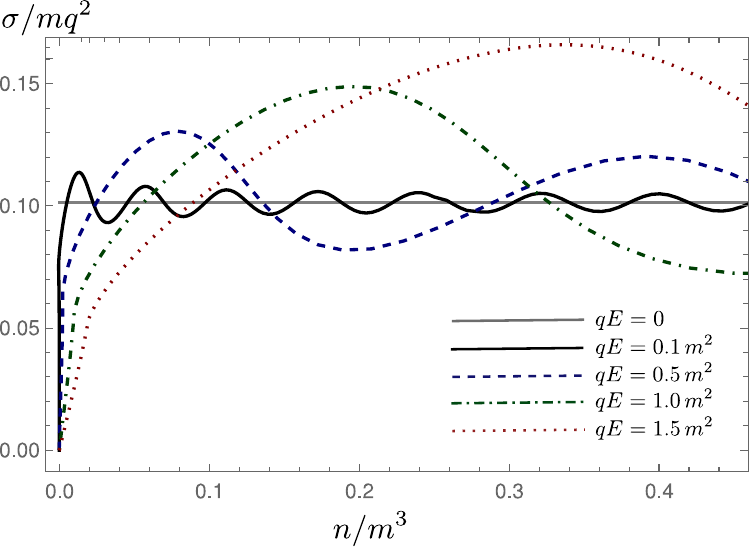}
 \caption{Electric conductivity normalized by $mq^2$ as a function of the number charge density in units of the fermion mass for different values of the external electric field.}
 \label{fig.Efixed}
\end{figure}

The electrical conductivity as a function of $n$ is plotted in Fig.\,\ref{fig.Efixed} for different values of the external electric field.
It is interesting to note the case where $E=0$ in which the conductivity becomes independent of $n$ (gray line).
When the electric field is turned on, the value of the conductivity grows rapidly from zero to $\sigma_0$ and then oscillates around this value.
As the external electric field increases, the wider the oscillations.

\bigskip


Observing the behavior of the conductivity for $E\neq 0$, it can be seen that all the curves start from $\sigma=0$ when $n=0$.
Therefore, it can be deduced that the case with $n=0$ and $E=0$ the conductivity must be zero.
In other words, the conductivity jumps from 0 to $\sigma_0$ as soon as a charge distribution appears in the system.

It is important to emphasize that quantities derived from time-ordered propagators are often associated with long-lived excitations or equilibrium properties of the system. This contrasts with retarded propagators or correlators, such as those appearing in the Kubo formula, which describe the causal, dissipative response of a system to external perturbations and are inherently tied to non-equilibrium or transient phenomena.

It is also important to remark that the aim here is to explore what a well-defined time-ordered propagator should be in the Schwinger proper-time formalism in the presence of an electric field, and to examine its consequences. In this sense, it is not applicable when performing calculations of the thermodynamic potential through the summation of Matsubara frequencies.

\section{Conclusions}

In this paper, the electric current density is obtained by means of a time-ordered Schwinger propagator for fermions immersed in a constant and uniform external electric field.
The time-ordered propagator can be well defined if we consider a time-independent electromagnetic vector potential.

In this scheme, the Schwinger phase  is considered as a non-local effective chemical potential term corresponding to the Coulomb potential, which depends on the mean position of the fermion correlator.
The effective chemical potential is an adjustable parameter that depends on the value of the number charge density and electric field.

This interpretation makes sense if one considers that the Coulomb potential for a constant electric field is ambiguous, and only the potential difference has physical meaning since it depends on the reference frame.
These types of interpretations are necessary because we are considering an electric field that permeates the entire space, so that no source of the field can be correctly identified or located.

The electrical conductivity obtained shows oscillations for electric field values beyond the leading order approximation.
This oscillation occurs around the conductivity near $E\sim 0$.
The electrical conductivity in the limit $E\to 0$ becomes independent of the number density, but vanishes in the absence of net charges.
Since $n$ corresponds to the number of fermions minus the number of antifermions in the system, this behavior should be independent of the pair creation predicted by Schwinger's theory for a strong electric field.

A comparison between this proposal and the formalism based on in- and out-state expectation values becomes necessary \cite{Copinger:2018ftr,Cao:2019hku}.
The expectation value for operators with external electric and magnetic field at finite temperature will be reported elsewhere.

\subsection*{Acknowledgments}
This article was funded by ANID/Fondecyt under grant No. 1250206.

\bibliographystyle{rmf-style}
\bibliography{Efield}

@article{Schwinger:1951nm,
    author = "Schwinger, Julian S.",
    editor = "Milton, K. A.",
    title = "{On gauge invariance and vacuum polarization}",
    doi = "https://doi.org/10.1103/PhysRev.82.664",
    journal = "Phys. Rev.",
    volume = "82",
    pages = "664--679",
    year = "1951"
}

@article{Chyi:1999fc,
    author = "Chyi, Tzuu-Kang and Hwang, Chien-Wen and Kao, W. F. and Lin, Guey-Lin and Ng, Kin-Wang and Tseng, Jie-Jun",
    title = "{The weak field expansion for processes in a homogeneous background magnetic field}",
    eprint = "hep-th/9912134",
    archivePrefix = "arXiv",
    doi = "https://doi.org/10.1103/PhysRevD.62.105014",
    journal = "Phys. Rev. D",
    volume = "62",
    pages = "105014",
    year = "2000"
}

@book{Kharzeev:2013jha,
    editor = "Kharzeev, Dmitri and Landsteiner, Karl and Schmitt, Andreas and Yee, H",
    title = "{Strongly Interacting Matter in Magnetic Fields}",
    year = "2013",
    doi = "https://doi.org/10.1007/978-3-642-37305-3",
    isbn = "978-3-642-37304-6, 978-3-642-37305-3",
    publisher = "Springer"
}

@article{Sauter:1931zz,
    author = "Sauter, Fritz",
    title = "{Uber das Verhalten eines Elektrons im homogenen elektrischen Feld nach der relativistischen Theorie Diracs}",
    doi = "https://doi.org/10.1007/BF01339461",
    journal = "Z. Phys.",
    volume = "69",
    pages = "742--764",
    year = "1931"
}

@article{Bzdak:2011yy,
    author = "Bzdak, Adam and Skokov, Vladimir",
    title = "{Event-by-event fluctuations of magnetic and electric fields in heavy ion collisions}",
    eprint = "1111.1949",
    archivePrefix = "arXiv",
    primaryClass = "hep-ph",
    reportNumber = "BNL-96541-2011-JA, RBRC-927",
    doi = "https://doi.org/10.1016/j.physletb.2012.02.065",
    journal = "Phys. Lett. B",
    volume = "710",
    pages = "171--174",
    year = "2012"
}

@article{Deng:2012pc,
    author = "Deng, Wei-Tian and Huang, Xu-Guang",
    title = "{Event-by-event generation of electromagnetic fields in heavy-ion collisions}",
    eprint = "1201.5108",
    archivePrefix = "arXiv",
    primaryClass = "nucl-th",
    doi = "https://doi.org/10.1103/PhysRevC.85.044907",
    journal = "Phys. Rev. C",
    volume = "85",
    pages = "044907",
    year = "2012"
}

@article{Tavares:2018poq,
    author = "Tavares, William R. and Avancini, Sidney S.",
    title = "{Schwinger mechanism in the SU(3) Nambu\textendash{}Jona-Lasinio model with an electric field}",
    eprint = "1801.10566",
    archivePrefix = "arXiv",
    primaryClass = "hep-ph",
    doi = "https://doi.org/10.1103/PhysRevD.97.094001",
    journal = "Phys. Rev. D",
    volume = "97",
    number = "9",
    pages = "094001",
    year = "2018"
}

@article{Loewe:2021mdo,
    author = "Loewe, M. and Valenzuela, D. and Zamora, R.",
    title = "{Catalysis and inverse electric catalysis in a scalar theory}",
    eprint = "2112.13872",
    archivePrefix = "arXiv",
    primaryClass = "hep-ph",
    doi = "https://doi.org/10.1103/PhysRevD.105.036017",
    journal = "Phys. Rev. D",
    volume = "105",
    number = "3",
    pages = "036017",
    year = "2022"
}

@article{Loewe:2022aaw,
    author = "Loewe, M. and Zamora, R.",
    title = "{Renormalons in a scalar self-interacting theory: Thermal, thermomagnetic, and thermoelectric corrections for all values of the temperature}",
    eprint = "2202.08873",
    archivePrefix = "arXiv",
    primaryClass = "hep-ph",
    doi = "https://doi.org/10.1103/PhysRevD.105.076011",
    journal = "Phys. Rev. D",
    volume = "105",
    number = "7",
    pages = "076011",
    year = "2022"
}

@article{Endrodi:2022wym,
    author = "Endr\H{o}di, Gergely and Mark\'o, Gergely",
    title = "{On electric fields in hot QCD: perturbation theory}",
    eprint = "2208.14306",
    archivePrefix = "arXiv",
    primaryClass = "hep-ph",
    doi = "https://doi.org/10.1007/JHEP12(2022)015",
    journal = "JHEP",
    volume = "12",
    pages = "015",
    year = "2022"
}

@article{Tavares:2023ybt,
    author = "Tavares, William R. and Avancini, Sidney S. and Farias, Ricardo L. S.",
    title = "{Quark matter under strong electric fields in the linear sigma model coupled with quarks}",
    eprint = "2305.07188",
    archivePrefix = "arXiv",
    primaryClass = "hep-ph",
    doi = "https://doi.org/10.1103/PhysRevD.108.016017",
    journal = "Phys. Rev. D",
    volume = "108",
    number = "1",
    pages = "016017",
    year = "2023"
}

@article{Cao:2015dya,
    author = "Cao, Gaoqing and Huang, Xu-Guang",
    title = "{Chiral phase transition and Schwinger mechanism in a pure electric field}",
    eprint = "1510.05125",
    archivePrefix = "arXiv",
    primaryClass = "nucl-th",
    doi = "https://doi.org/10.1103/PhysRevD.93.016007",
    journal = "Phys. Rev. D",
    volume = "93",
    number = "1",
    pages = "016007",
    year = "2016"
}

@article{Tavares:2019mvq,
    author = "Tavares, William R. and Farias, Ricardo L. S. and Avancini, Sidney S.",
    title = "{Deconfinement and chiral phase transitions in quark matter with a strong electric field}",
    eprint = "1912.00305",
    archivePrefix = "arXiv",
    primaryClass = "hep-ph",
    doi = "https://doi.org/10.1103/PhysRevD.101.016017",
    journal = "Phys. Rev. D",
    volume = "101",
    number = "1",
    pages = "016017",
    year = "2020"
}

@article{Ahmad:2020ifp,
    author = "Ahmad, Aftab",
    title = "{Chiral symmetry restoration and deconfinement in the contact interaction model of quarks with parallel electric and magnetic fields}",
    eprint = "2009.09482",
    archivePrefix = "arXiv",
    primaryClass = "hep-ph",
    doi = "https://doi.org/10.1088/1674-1137/abfb5f",
    journal = "Chin. Phys. C",
    volume = "45",
    number = "7",
    pages = "073109",
    year = "2021"
}

@article{Ahmad:2023mqg,
    author = "Ahmad, Aftab and Farooq, Akif",
    title = "{Schwinger Pair Production in QCD from Flavor-Dependent Contact Interaction Model of Quarks}",
    eprint = "2302.13265",
    archivePrefix = "arXiv",
    primaryClass = "hep-ph",
    doi = "https://doi.org/10.1007/s13538-024-01581-0",
    journal = "Braz. J. Phys.",
    volume = "54",
    number = "6",
    pages = "212",
    year = "2024"
}

@article{Loewe:2022plp,
    author = "Loewe, M. and Valenzuela, D. and Zamora, R.",
    title = "{Effective potential and mass behavior of a self-interacting scalar field theory due to thermal and external electric and magnetic fields effects}",
    eprint = "2207.12387",
    archivePrefix = "arXiv",
    primaryClass = "hep-ph",
    doi = "https://doi.org/10.1140/epja/s10050-023-01097-2",
    journal = "Eur. Phys. J. A",
    volume = "59",
    number = "8",
    pages = "184",
    year = "2023"
}

@article{PhysRevD.91.025011,
  title = {Chiral Hall effect and chiral electric waves},
  author = {Pu, Shi and Wu, Shang-Yu and Yang, Di-Lun},
  journal = {Phys. Rev. D},
  volume = {91},
  issue = {2},
  pages = {025011},
  numpages = {24},
  year = {2015},
  month = {Jan},
  publisher = {American Physical Society},
  doi = {10.1103/PhysRevD.91.025011},
}

@article{PhysRevD.91.045001,
  title = {Chiral electric separation effect in the quark-gluon plasma},
  author = {Jiang, Yin and Huang, Xu-Guang and Liao, Jinfeng},
  journal = {Phys. Rev. D},
  volume = {91},
  issue = {4},
  pages = {045001},
  numpages = {9},
  year = {2015},
  month = {Feb},
  publisher = {American Physical Society},
  doi = {10.1103/PhysRevD.91.045001},
}

@article{Huang:2015oca,
    author = "Huang, Xu-Guang",
    title = "{Electromagnetic fields and anomalous transports in heavy-ion collisions --- A pedagogical review}",
    eprint = "1509.04073",
    archivePrefix = "arXiv",
    primaryClass = "nucl-th",
    doi = "https://doi.org/10.1088/0034-4885/79/7/076302",
    journal = "Rept. Prog. Phys.",
    volume = "79",
    number = "7",
    pages = "076302",
    year = "2016"
}

@article{Copinger:2018ftr,
    author = "Copinger, Patrick and Fukushima, Kenji and Pu, Shi",
    title = "{Axial Ward identity and the Schwinger mechanism -- Applications to the real-time chiral magnetic effect and condensates}",
    eprint = "1807.04416",
    archivePrefix = "arXiv",
    primaryClass = "hep-th",
    doi = "https://doi.org/10.1103/PhysRevLett.121.261602",
    journal = "Phys. Rev. Lett.",
    volume = "121",
    number = "26",
    pages = "261602",
    year = "2018"
}

@article{Cao:2019hku,
    author = "Cao, Gaoqing",
    title = "{The electromagnetic field effects in in-out and in-in formalisms}",
    eprint = "1912.11971",
    archivePrefix = "arXiv",
    primaryClass = "nucl-th",
    doi = "https://doi.org/10.1016/j.physletb.2020.135477",
    journal = "Phys. Lett. B",
    volume = "806",
    pages = "135477",
    year = "2020"
}

@article{Murguia:2009fs,
    author = "Murguia, Gabriela and Raya, Alfredo and Sanchez, Angel and Reyes, Edward",
    title = "{The Electron Propagator in External Electromagnetic Fields in Lower Dimensions}",
    eprint = "0910.1881",
    archivePrefix = "arXiv",
    primaryClass = "hep-th",
    doi = "https://doi.org/10.1119/1.3311656",
    journal = "Am. J. Phys.",
    volume = "78",
    pages = "700--707",
    year = "2010"
}

@article{Ayala:2010fm,
    author = "Ayala, Alejandro and Bashir, Adnan and Gutierrez, Enif and Raya, Alfredo and Sanchez, Angel",
    title = "{Chiral and Parity Symmetry Breaking for Planar Fermions: Effects of a Heat Bath and Uniform External Magnetic Field}",
    eprint = "1007.4249",
    archivePrefix = "arXiv",
    primaryClass = "hep-ph",
    doi = "https://doi.org/10.1103/PhysRevD.82.056011",
    journal = "Phys. Rev. D",
    volume = "82",
    pages = "056011",
    year = "2010"
}

@article{Dudal:2021ret,
    author = "Dudal, David and Matusalem, Filipe and Mizher, Ana J\'ulia and Rocha, Alexandre Reily and Villavicencio, Cristian",
    title = "{Half-integer anomalous currents in 2D materials from a QFT viewpoint}",
    eprint = "2103.10341",
    archivePrefix = "arXiv",
    primaryClass = "cond-mat.mes-hall",
    doi = "https://doi.org/10.1038/s41598-022-09483-4",
    journal = "Sci. Rep.",
    volume = "12",
    number = "1",
    pages = "5439",
    year = "2022"
}

@article{Tsamis:2000ah,
    author = "Tsamis, N. C. and Woodard, R. P.",
    title = "{Schwinger's propagator is only a Green's function}",
    eprint = "hep-ph/0007167",
    archivePrefix = "arXiv",
    reportNumber = "CRETE-00-12, UFIFT-HEP-00-12",
    doi = "https://doi.org/10.1088/0264-9381/18/1/306",
    journal = "Class. Quant. Grav.",
    volume = "18",
    pages = "83--94",
    year = "2001"
}

@article{Shuryak:1980tp,
    author = "Shuryak, Edward V.",
    title = "{Quantum Chromodynamics and the Theory of Superdense Matter}",
    doi = "https://doi.org/10.1016/0370-1573(80)90105-2",
    journal = "Phys. Rept.",
    volume = "61",
    pages = "71--158",
    year = "1980"
}

@article{PhysRevD.42.2881,
  title = {QED with a chemical potential: The case of a constant magnetic field},
  author = {Chodos, Alan and Everding, Kenneth and Owen, David A.},
  journal = {Phys. Rev. D},
  volume = {42},
  issue = {8},
  pages = {2881--2892},
  numpages = {0},
  year = {1990},
  month = {Oct},
  publisher = {American Physical Society},
  doi = {https://doi.org/10.1103/PhysRevD.42.2881},
}

@article{Mizher:2013kza,
    author = "Mizher, Ana Julia and Raya, Alfredo and Villavicencio, Cristian",
    title = "{Electric current generation in distorted graphene}",
    eprint = "1312.3274",
    archivePrefix = "arXiv",
    primaryClass = "hep-ph",
    doi = "https://doi.org/10.1142/S0217979215502574",
    journal = "Int. J. Mod. Phys. B",
    volume = "30",
    number = "2",
    pages = "1550257",
    year = "2015"
}

@article{Mizher:2018dtf,
    author = "Mizher, Ana Julia and Hernandez-Ortiz, Saul and Raya, Alfredo and Villavicencio, Cristian",
    title = "{Aspects of the pseudo Chiral Magnetic Effect in 2D Weyl-Dirac Matter}",
    eprint = "1803.05794",
    archivePrefix = "arXiv",
    primaryClass = "hep-ph",
    doi = "https://doi.org/10.1140/epjc/s10052-018-6380-1",
    journal = "Eur. Phys. J. C",
    volume = "78",
    number = "11",
    pages = "912",
    year = "2018"
}

@book{schwartz2014quantum,
  title={Quantum Field Theory and the Standard Model},
  author={Schwartz, M.D.},
  isbn={9781107034730},
  lccn={2013016195},
  series={Quantum Field Theory and the Standard Model},
  url={https://books.google.cl/books?id=HbdEAgAAQBAJ},
  year={2014},
  publisher={Cambridge University Press}
}

@book{Dittrich:2000zu,
    author = "Dittrich, W. and Gies, H.",
    title = "{Probing the quantum vacuum. Perturbative effective action approach in quantum electrodynamics and its application}",
    doi = "https://doi.org/10.1007/3-540-45585-X",
    isbn = "978-3-540-67428-3, 978-3-540-45585-1",
    volume = "166",
    year = "2000",
    publisher="{Springer}"
}

\end{multicols}
\end{document}